\documentclass[pre,twocolumn,showpacs,floatfix,amsmath,amssymb]{revtex4-1}
\usepackage{graphicx}
\usepackage{dcolumn}
\usepackage{bm}
\usepackage{color}

\newcommand{\figOne}{
\begin{figure}[!ht]
    \centering
        \includegraphics[width=3.5in]{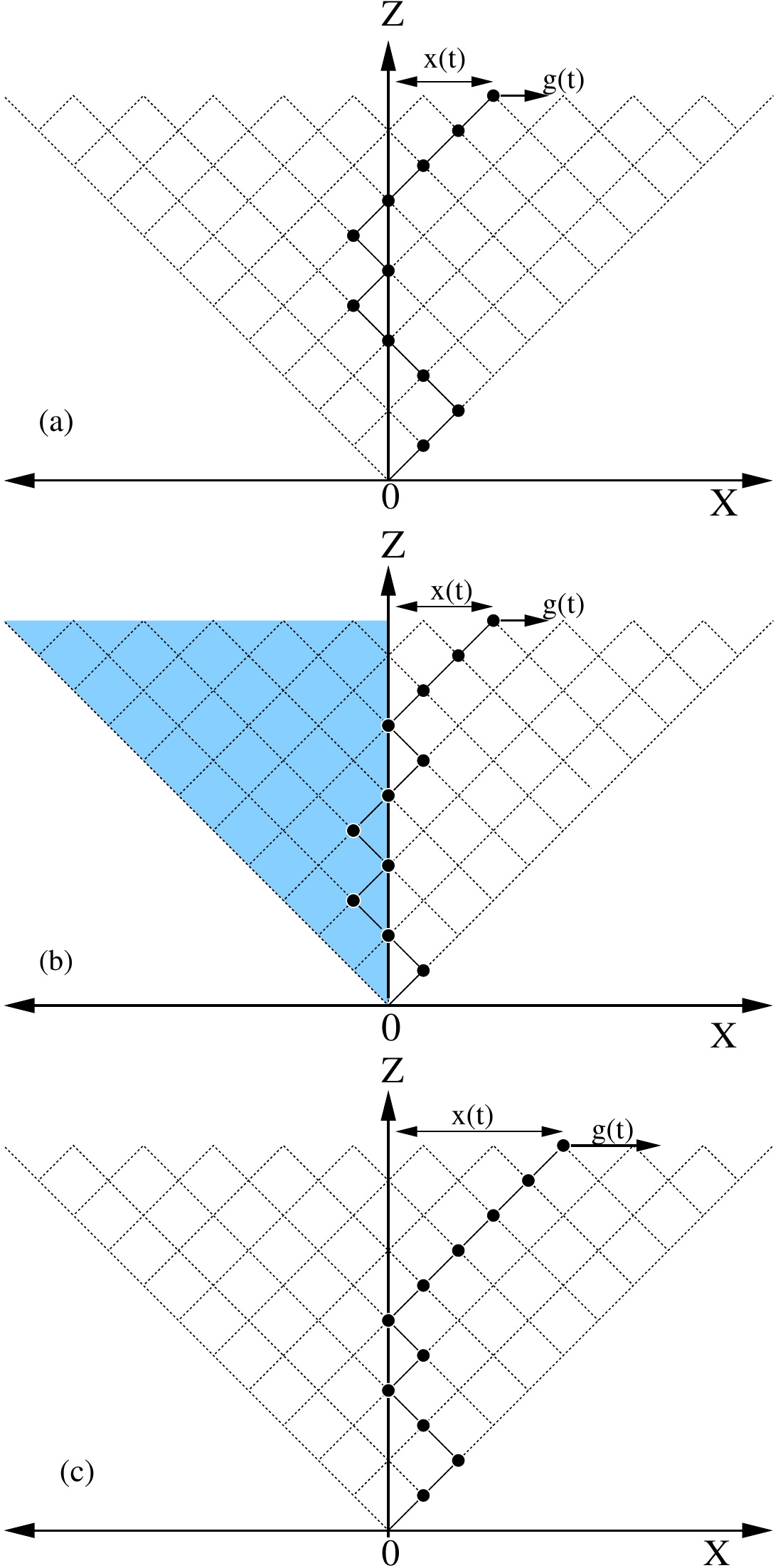}
	\caption {Schematic diagram of a directed polymer adsorbed on a surface (along $z$-axis) at $x = 0$. (a) Soft-wall: polymer is allowed in the whole region. (b) Wall separating two different types of media: polymer is allowed in whole region, however there is an extra repulsive potential $V(> 0)$ on side $x < 0$. (c) Hard-wall: polymer is not allowed in the region $x < 0$. One end of the polymer is anchored at the origin ($O$), and the chain on the free end is subjected to a time-dependent periodic force with frequency
	$\omega$ and amplitude $g_0$. \label{fig:1}}
\end{figure}
}
\newcommand{\figTwo}{
\begin{figure*}[!t]
    \centering
        \includegraphics[width=0.9\linewidth]{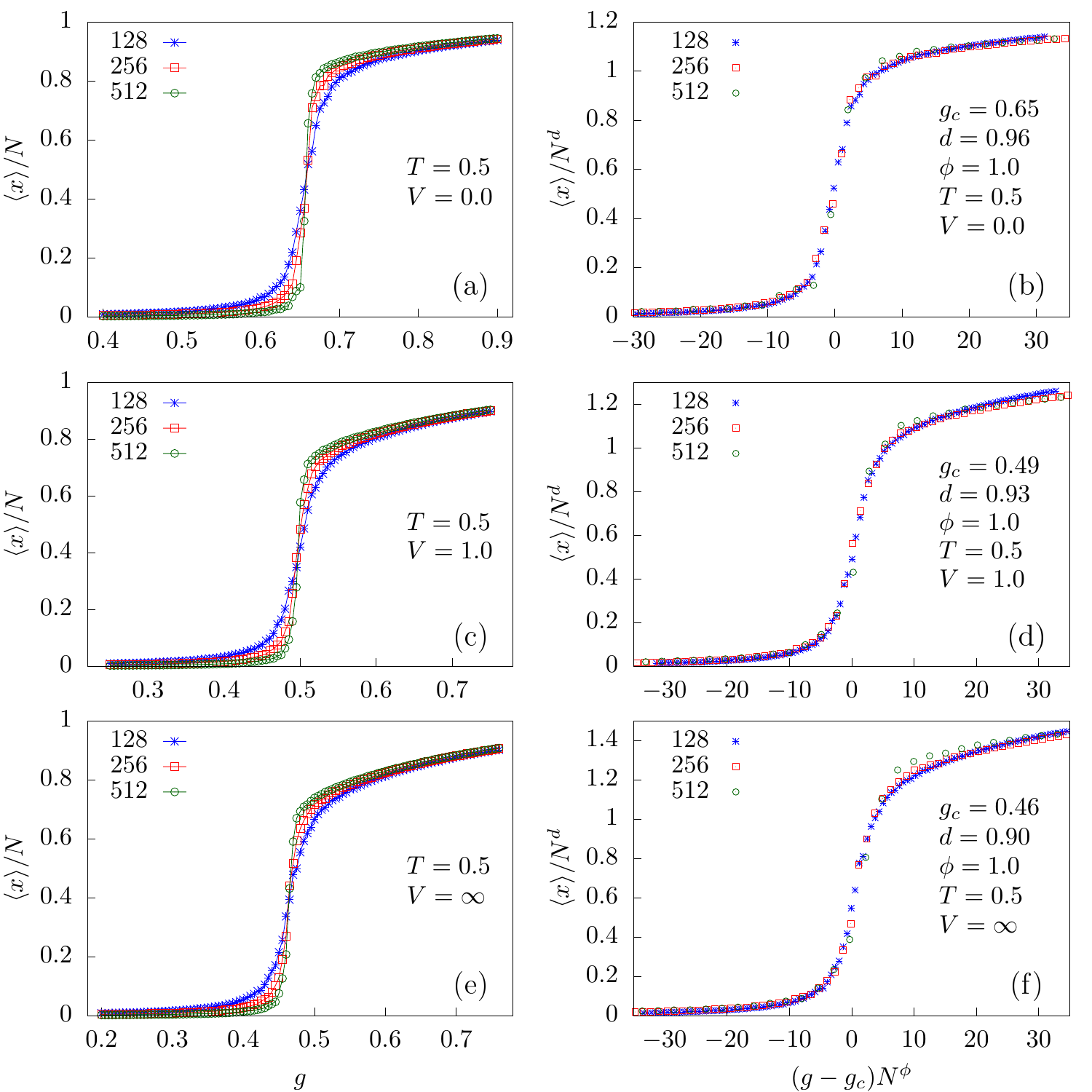}
	\caption {Scaled extension $\langle x \rangle / N$, as a function of constant pulling force $g$, obtained using Monte Carlo simulations, for different chain lengths $N = 128, 256$, and $512$ at $T = 0.5$ for (a) soft-wall. (b) $\langle x \rangle / N^d$ as a function of $(g-g_c)N^\phi$ showing a nice collapse of data for $g_c = 0.65\pm0.02$, $d = 0.96 \pm 0.05$, and $\phi = 1.0\pm0.02$. (c) For wall separating two different types of media. (d) Data collapse of (c) for $g_c = 0.49\pm0.02$, $d = 0.93 \pm 0.10$, and $\phi = 1.0\pm0.02$. (e) For hard-wall case. (f) Collapse of data shown in (e) for $g_c = 0.46\pm0.02$, $d = 0.90 \pm 0.10$, and $\phi = 1.0\pm0.02$.\label{fig:2}}
\end{figure*}
}

\newcommand{\figThree}{
\begin{figure}[!t]
    \centering
        \includegraphics[width=3.5in]{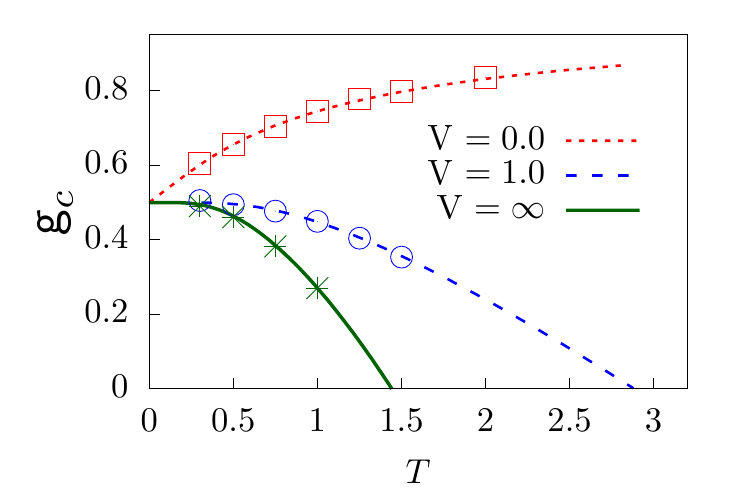}
	\caption {Critical unzipping force $g_c$ as a function of temperature $T$ for soft-wall ($V = 0.0$), wall separating two different types of media ($V = 1.0$), and hard-wall ($V = \infty$). The lines are
	the exact results obtained from the generating function approach, and the points are obtained by using finite-size scaling of the force-distance isotherms as obtained from the Monte Carlo simulations. \label{fig:3}}
\end{figure}
}

\newcommand{\figFour}{
\begin{figure*}[!t]
    \centering
        \includegraphics[width=0.9\linewidth]{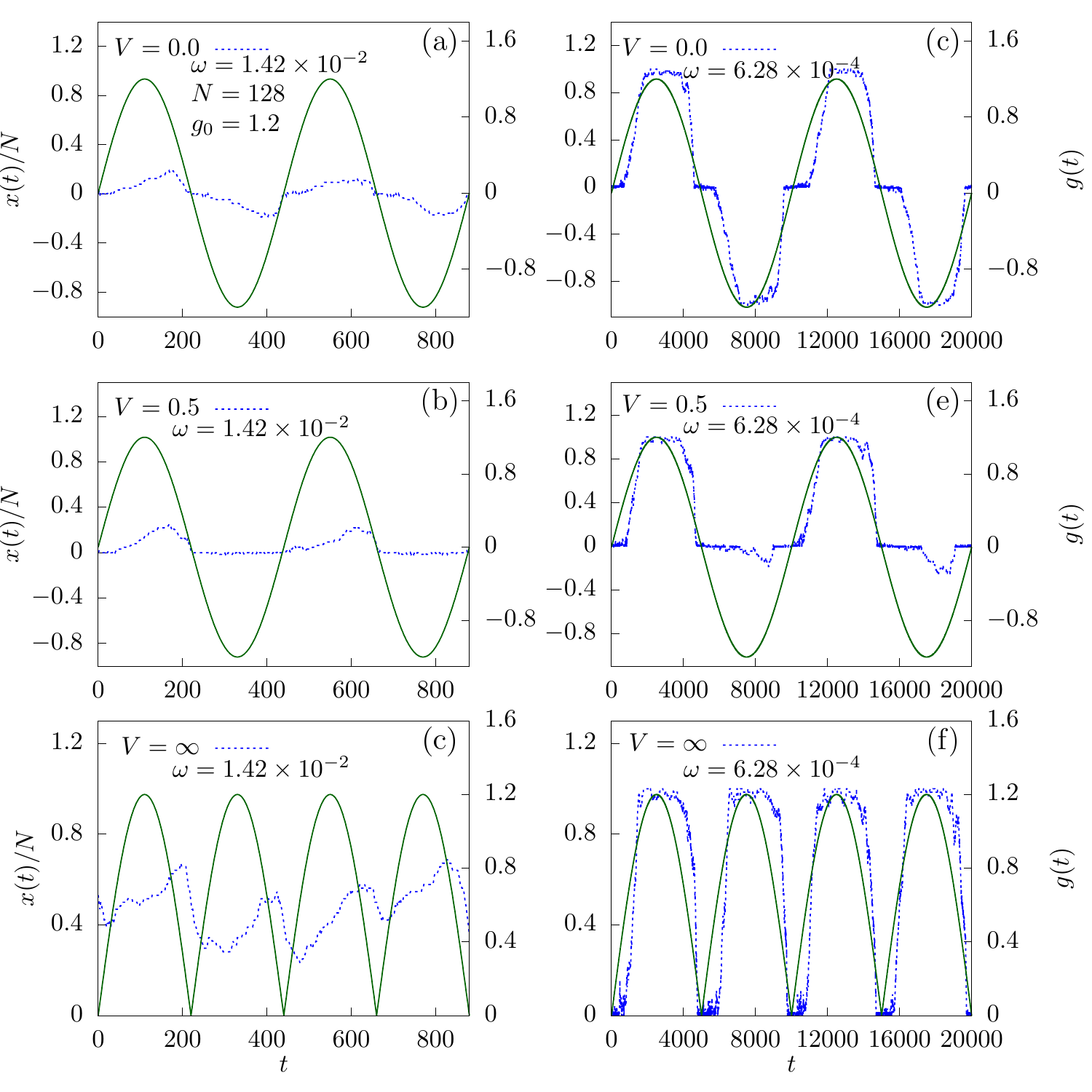}
	\caption {The scaled extension $x(t)/N$ from the wall of end monomer of the polymer of length $N = 128$ as a function
	of time $t$ when it is subjected to a periodic force of amplitude
	$g_0 = 1.2$ at frequency $\omega = 1.42\times 10^{-2}$. For the
    (a) soft-wall ($V = 0.0$), (b) wall separating two different media ($V = 0.5$), and (c) hard-wall ($V = \infty $). Plots (d), (e), and
	(f) are same as plots (a), (b), and (c) at frequency $\omega =
	6.28\times 10^{-4}$. The time variation of force, $g(t)$, is
	represented by solid lines. \label{fig:4}}
\end{figure*}
}

\newcommand{\figFive}{
\begin{figure*}[!t]
    \centering
        \includegraphics[width=0.42\linewidth]{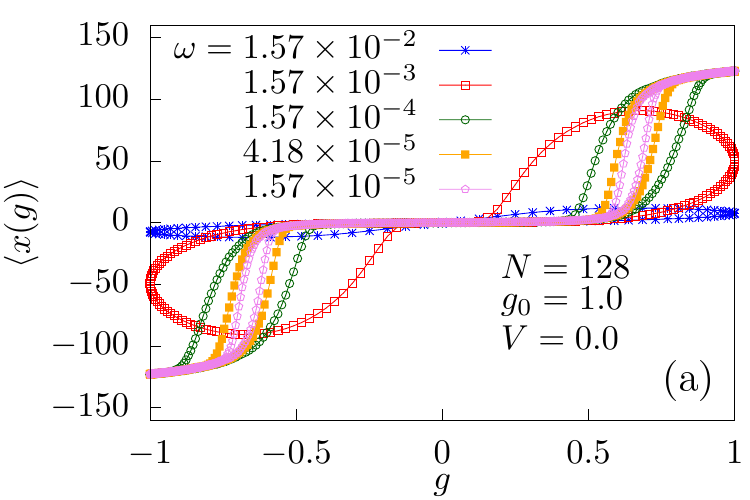}
        \includegraphics[width=0.42\linewidth]{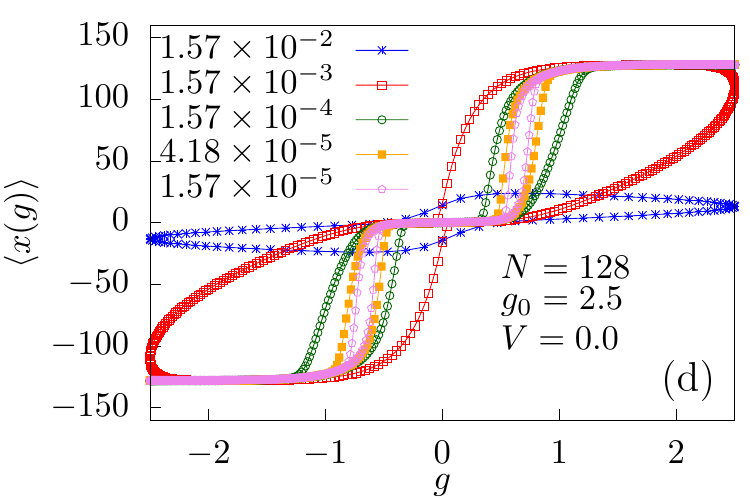}
        \includegraphics[width=0.42\linewidth]{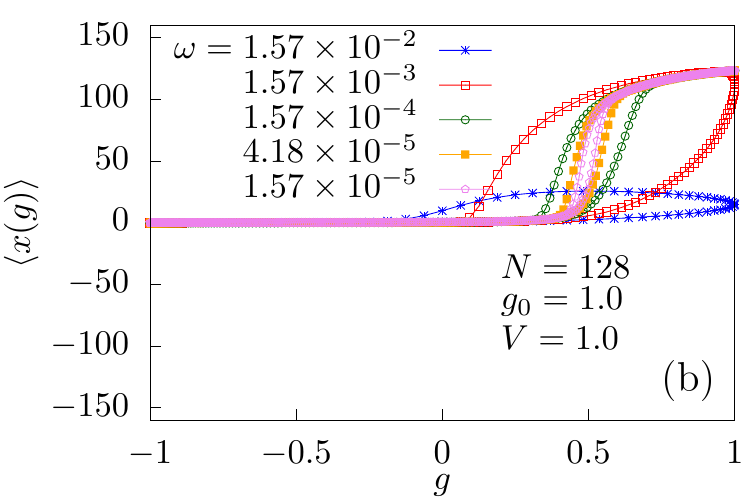}
        \includegraphics[width=0.42\linewidth]{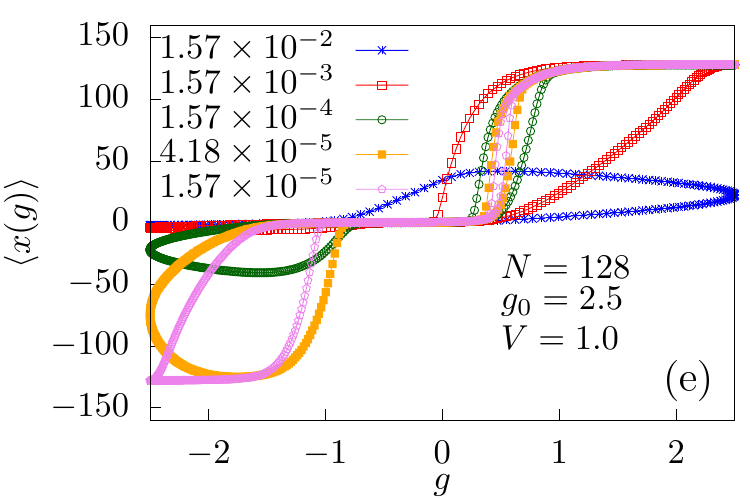}
        \includegraphics[width=0.42\linewidth]{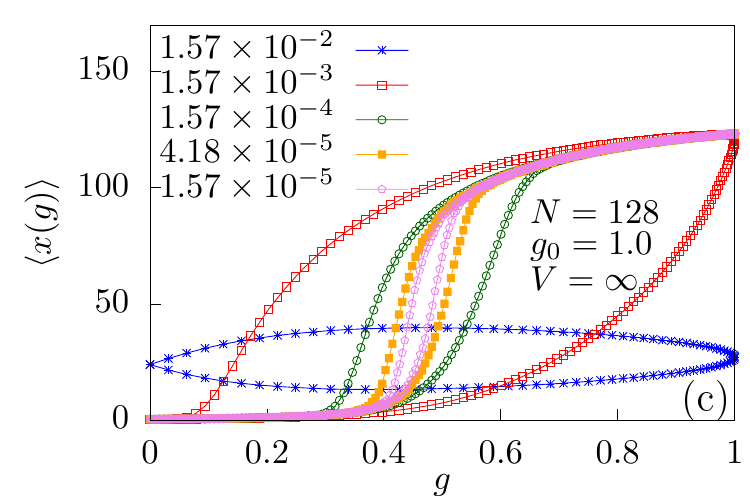}
        \includegraphics[width=0.42\linewidth]{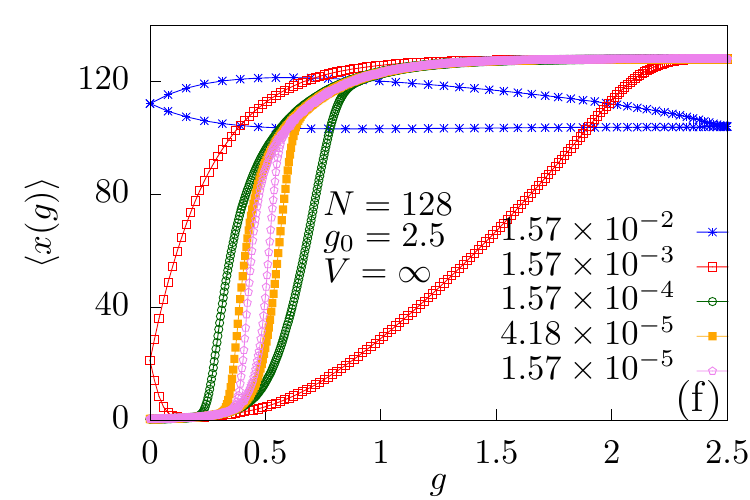}
	    \caption {The force $g$ vs extension $\langle x(g) \rangle$ curves averaged over $10^4$ cycles of a polymer of length $N = 128$ for the various frequencies for (a) soft-wall, (b) wall separating two different type of media, and (c) hard-wall at force amplitudes $g_0 = 1.0$. Plots (d), (e) and (f) are the same as the plots (a), (b) and (c) respectively, at $g_0 = 2.5$. The line joining the points in these plots is just a guide for the eye. \label{fig:5}}
\end{figure*}
}

\newcommand{\figSix}{
\begin{figure*}[!t]
    \centering
    \includegraphics[width=0.85\linewidth]{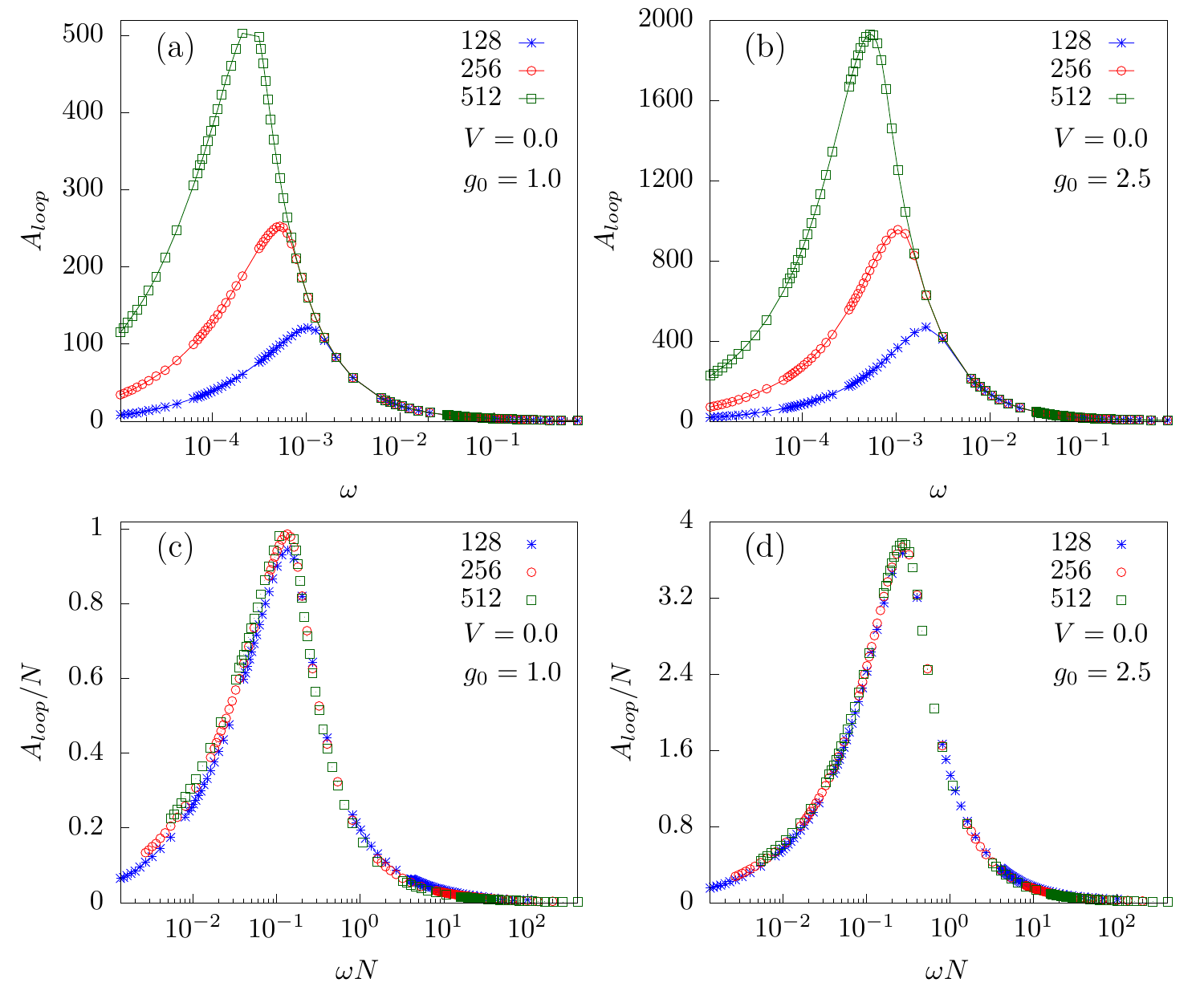}
     \caption {Area of the hysteresis loop $A_{loop}$ as a function of frequency $\omega$, in a semilog scale, for the polymer of lengths $N = 128, 256$, and $512$ at force amplitudes (a) $g_0 = 1.0$ and (b) $g_0 = 2.5$ when it is adsorbed on a soft-wall $(V = 0.0)$. Plots (c) and (d) are $A_{loop}/N$ vs $\omega N$ of the data shown in (a) and (b), respectively. The line joining the points in these plots is just a guide for the eye. \label{fig:6}}
\end{figure*}
}

\newcommand{\figSeven}{
\begin{figure*}[!t]
    \centering
     \includegraphics[width=0.85\linewidth]{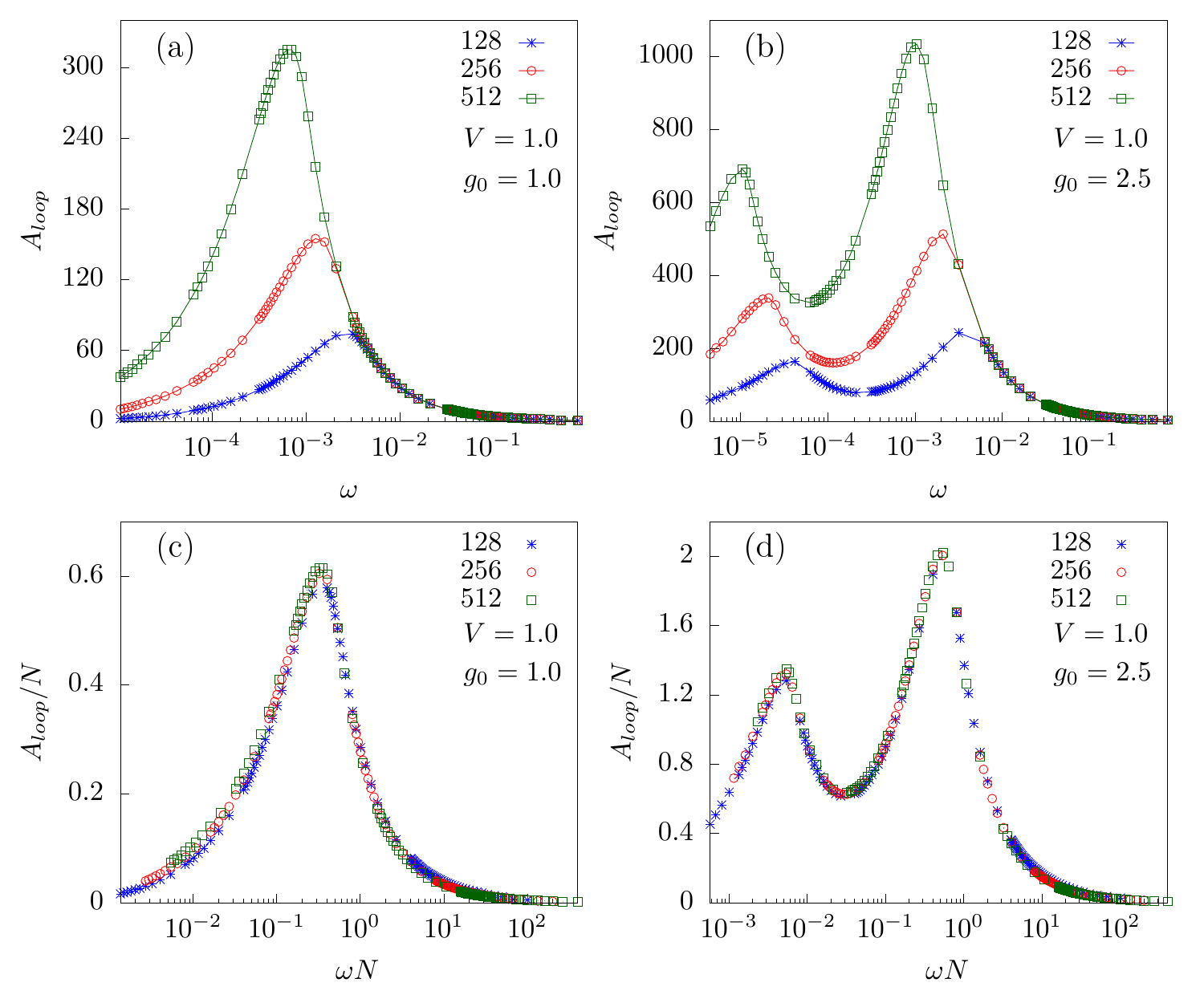}
     \caption {Area of the hysteresis loop $A_{loop}$ as a function of frequency $\omega$, in a semilog scale, for the polymer of lengths $N = 128, 256$, and $512$ at force amplitudes (a) $g_0 = 1.0$ and (b) $g_0 = 2.5$ when it is adsorbed on wall separating two different media $(V = 1.0)$. Plots (c) and (d) are $A_{loop}/N$ vs $\omega N$ of the data shown in (a) and (b), respectively. The line joining the points in these plots is just a guide for the eye. \label{fig:7}}
\end{figure*}
}

\newcommand{\figEight}{
\begin{figure*}[!t]
    \centering
     \includegraphics[width=0.85\linewidth]{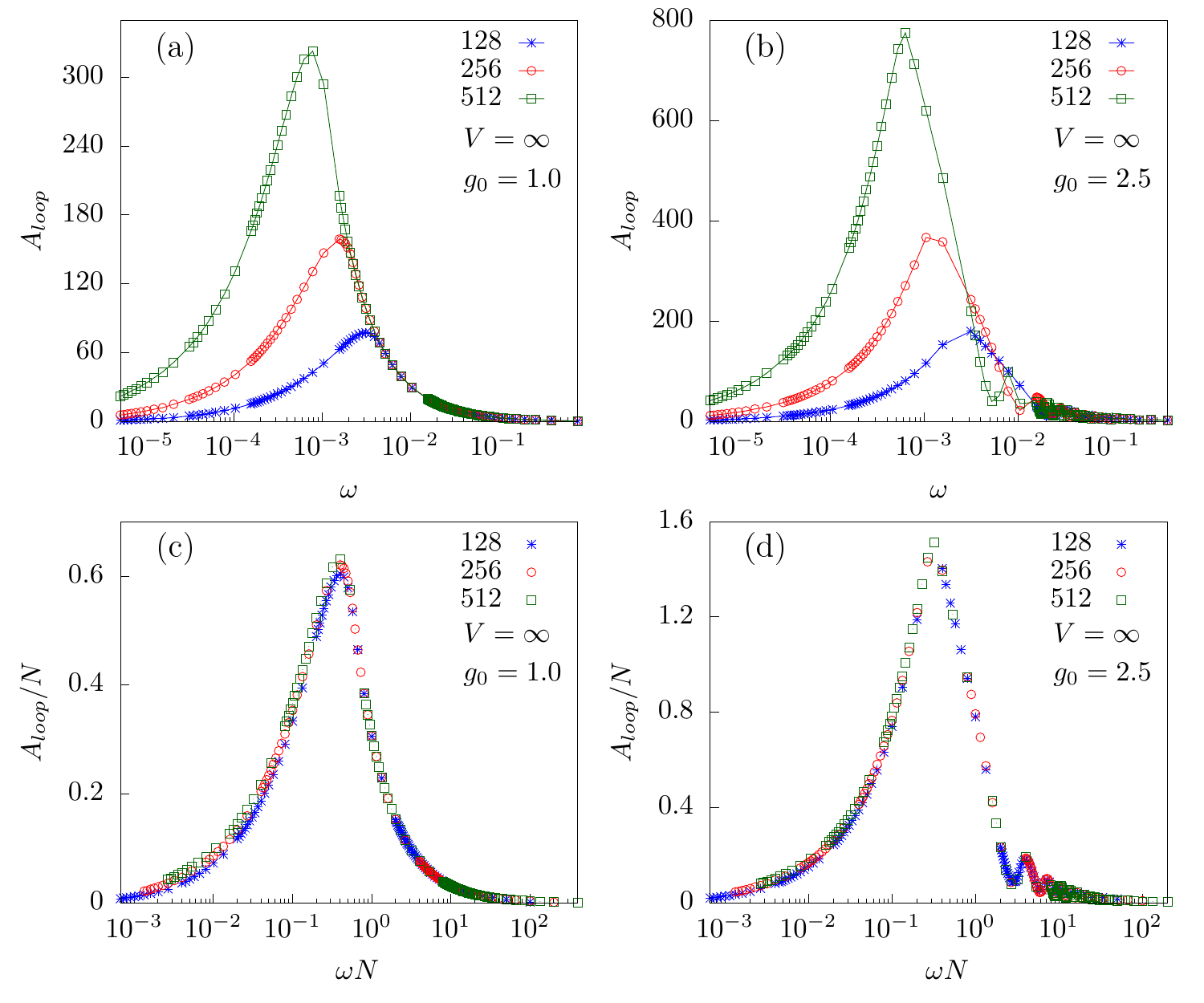}
     \caption {Area of the hysteresis loop $A_{loop}$ as a function of frequency $\omega$, in a semilog scale, for the polymer of lengths $N = 128, 256$, and $512$ at force amplitudes (a) $g_0 = 1.0$ and (b) $g_0 = 2.5$ when it is adsorbed on a hard-wall $(V = \infty)$. Plots (c) and (d) are $A_{loop}/N$ vs $\omega N$ of the data shown in (a) and (b), respectively. The line joining the points in these plots is just a guide for the eye. \label{fig:8}}
\end{figure*}
}
\newcommand{\figNine}{
\begin{figure*}[!t]
    \centering
     \includegraphics[width=0.95\linewidth]{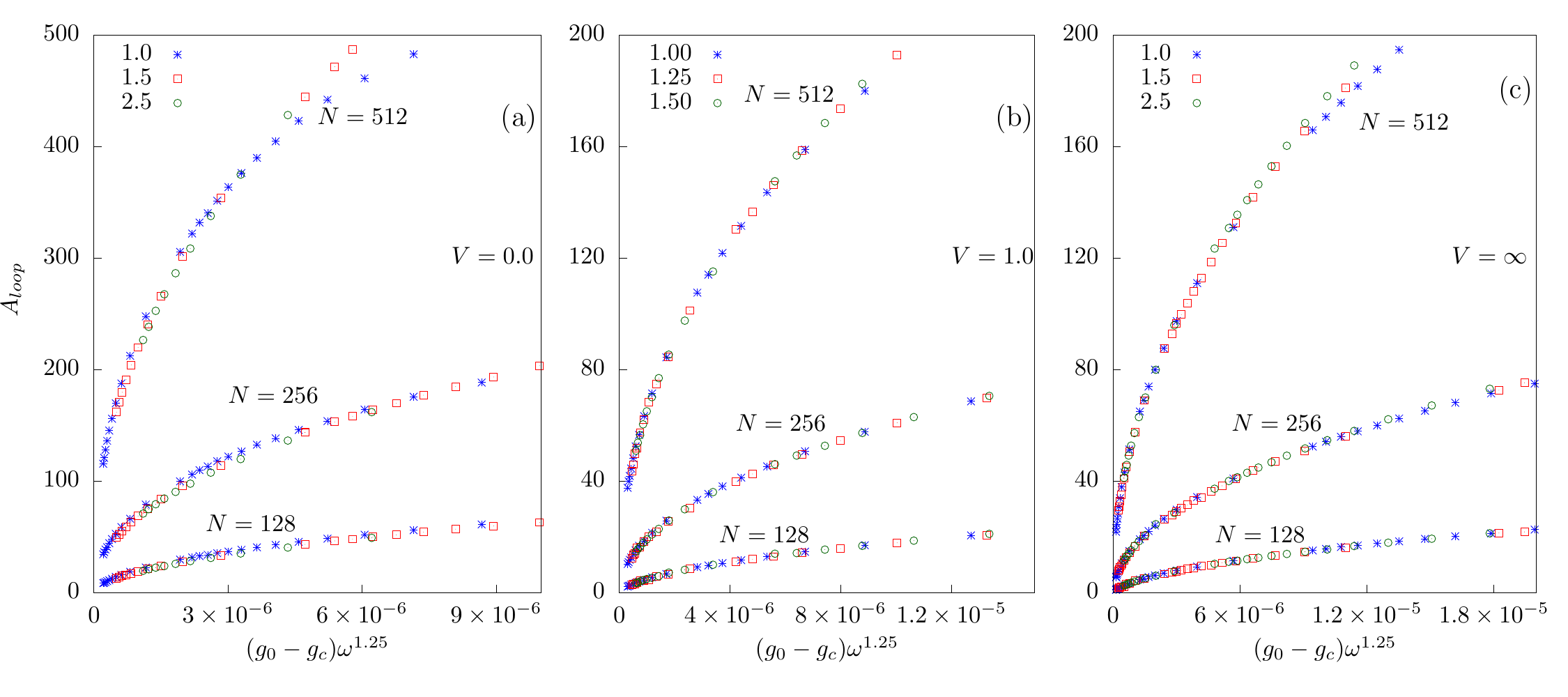}
     \caption {$A_{loop}$ as a function of $(g_0 - g_c) \omega^{1.25}$ for the polymer of lengths $N = 128, 256$, and $512$ at various force amplitudes for (a) soft-wall $(V = 0.0 )$, (b) wall separating two different media $(V = 1.0 )$, and (c) hard-wall $(V = \infty)$. \label{fig:9}}
\end{figure*}
}

\begin{document}
\title{Study of unzipping transitions in an adsorbed polymer by a periodic force }
\author{Ramu Kumar Yadav}
\email{ramukumar@iisermohali.ac.in}
\affiliation{\small \it Department of Physical Sciences, Indian Institute of Science Education and Research Mohali, Sector 81, Knowledge City, S. A. S. Nagar, Manauli PO 140306, India}
\date{\today}

\begin{abstract}

Using Monte Carlo simulations, we study the dynamic transitions in the unzipping of an adsorbed homogeneous polymer on a surface (or wall). We consider three different types of surfaces. One end of the polymer is always kept anchored, and other end monomer is subjected to a periodic force with frequency $\omega$ and amplitude $g_0$. We observe that the force-distance isotherms show hysteresis loops in all the three cases. 
For all the three cases, it is found that the area of the hysteresis loop, $A_{loop}$, scales as $1/\omega$ in the higher frequency regime, and as $g_0^{\alpha} \omega^{\beta}$ with exponents $\alpha = 1$ and $\beta = 1.25$ in the lower frequency regime. The values of exponents $\alpha$ and $\beta$ are similar to the exponents obtained in the earlier Monte Carlo simulation studies of DNA chains and a Langevin dynamics simulation
study of longer DNA chains.
\end{abstract}

\maketitle

\section{Introduction}

There have been many studies of the unzipping transitions of an adsorbed polymer from an attractive surface because it finds a wide variety of applications such as in lubrication, adhesion, surface protection, coating of surfaces, wetting, and many processes in physics and biology.
The development of the single-molecule manipulation techniques, which are used to study individual polymer chains and biological molecules by applying mechanical forces in the pico-newton ranges, has made it possible to understand the mechanical properties and the characterization of intermolecular interactions of these molecules~\cite{Strick2001,Celestini2004}.
A biological relevant problem, which belongs to the same universality class and has great importance in biological processes like DNA replication and RNA transcription~\cite{Watson2004}, is the unzipping of a double-stranded DNA (dsDNA) by an external force, which has been studied over two decades both theoretically~\cite{Bhattacharjee2000,Lubensky2000,Sebastian2000,Marenduzzo2001,Marenduzzo2002,Kapri2004,Kumar2010} and experimentally~\cite{Bockelmann2002,Danilowicz2003,Danilowicz2004,Ritort2006,Hatch2007}. The unzipping transition, in the presence of a static pulling force, is well studied and is a first-order phase transition. If a polymer adsorbed on a surface is subjected to an external force, at a given temperature $T$, which is below its desorption temperature $T_m$, the polymer remains in the adsorbed phase if the pulling force is less than a critical value, $g_c(T)$~\cite{Kapri2005,Orlandini2004,Iliev2004,Mishra2004,Bhattacharya2009}, which depends on the temperature. 
If the force exceeds $g_c(T)$, the polymer is in the unzipped phase. 
When the adsorbed polymer is subjected to a periodic force the force-distance isotherms show hysteresis, similar to the one seen in case of dsDNA. The polymer can be taken from the zipped state to an unzipped state, or \textit{vice versa}, dynamically either by varying the frequency keeping the amplitude constant or varying the amplitude keeping the frequency constant. The study of hysteresis in unbinding and rebinding of biomolecules and polymers is important because it provides useful information about the kinetics of conformational transformations, the potential energy landscape, and
controlling the folding pathway of a single molecule and in
force sensor studies~\cite{Hatch2007,Friddle2008,Tshiprut2009,Li2007,Yasunaga2019}.

Over the past one decade, the behavior of a dsDNA under a periodic force with frequency $\omega$ and amplitude $g_0$ has been studied by using Langevin dynamics simulation of an off-lattice coarse-grained model for short as well as long homo-polymer DNA chains~\cite{Kumar2013,Mishra2013,Mishra2013jcp,Kumar2016,Pal2018,Kapri2021}, and by using Monte Carlo simulations on a relatively longer chains of directed self-avoiding walk (DSAW) model of a homo-polymer dsDNA~\cite{Kapri2012,Kapri2014,Kalyan2019} and a block copolymer DNA~\cite{Yadav2021} on a 2-dimensional square lattice.

In this paper, we use Monte Carlo simulations to study the dynamic transitions in the unzipping of an adsorbed polymer on a surface (or wall) by a pulling force. We considered three different types of the walls. The one end of the polymer is under the influence of a pulling force while other end is kept anchored. We studied both the constant and the periodic pulling force cases. For a constant pulling force, the equilibrium phase boundary which separates the zipped and the unzipped phases is found to be the same as the phase boundary obtained analytically using the generating function technique~\cite{Kapri2005} in all the three cases of the wall. For the periodic force case, we observed that the force-distance isotherms show hysteresis loops. 
The behavior of the loop area, $A_{loop}$, which represents the energy dissipated in the system, depends on both the frequency and the amplitude of the force. We found that $A_{loop}$ shows nonmonotonic behavior as the frequency is varied keeping the amplitude constant. For all the three cases, we found that $A_{loop}$ scales as $1/\omega$ in the higher frequency regime, and as $g_0^{\alpha} \omega^{\beta}$ with exponents $\alpha = 1$ and $\beta = 1.25$ in the lower frequency regime. These exponents are similar to the exponents obtained in the earlier Monte Carlo simulation studies of DNA chains~\cite{Kapri2014,Yadav2021} and a Langevin dynamics simulation study of a longer DNA chain~\cite{Kapri2021} in 2-dimensions.

The paper is organized as follows: In Sec. \ref{sec:Model}, we describe the simulating model and the quantities of interest that we study in this paper. We discuss our results for both the static and the periodic
pulling force cases in Sec. \ref{sec:Results}. The results of this paper are summarized in Sec. \ref{sec:Summary}. 

\figOne

\section{Model} {\label{sec:Model} }
 
We model the polymer by a directed self-avoiding random walk in $d = 1 + 1$ dimensional square lattice. The walk starts at the origin $O$, and is restricted to go towards the positive direction of the diagonal axis ($z$ direction). The directional nature of the walks takes care of the self-avoidance. We put an attractive wall along $z$-axis at $x = 0$. Whenever a monomer is on the wall, polymer gains an energy $-\epsilon$($\epsilon > 0$).  One end of the polymer is always kept anchored to the wall at the origin, and other end monomer is subjected to a time-dependent periodic force $g(t)$, which acts along the transverse direction ($x$ direction) and is given by 
\begin{equation}
     g(t) = g_0 \sin(\omega t), 
     \label{eq:Twoforce}  
\end{equation}
where $g_0$ is the amplitude and $\omega$, is the frequency. 
we consider three different types of walls:
\begin{enumerate}
    
    \item In the first type of wall, the polymer is allowed to stay only on one side of the wall, i.e., the wall is impenetrable. We refer such a surface as \textit{hard-wall}. 
    
    \item  In the second type of wall, the polymer is allowed to cross the surface, and it has equal affinity on both the sides of the surface. We call such a surface as \textit{soft-wall}. 
    
    \item In the third type of wall, the polymer is allowed to cross the surface, but it has different affinities on either sides of the surface. This can be thought of as an interface between two immiscible liquids in which the polymer has different degree of solubility.  
\end{enumerate}  
The three different possibilities for the wall can be modeled by assigning a repulsive potential $V (> 0)$ on one side of the wall, say $x < 0$. In the two extreme limits, i.e., (i) for $V = 0$, the wall behaves as a soft-wall, and (ii) for $V = \infty$, it behaves like a hard-wall. All other values of $V$, the wall acts like an interface between the two immiscible liquids with different affinities with the polymer. The schematic diagram of the model is shown in Fig.~\ref{fig:1}.

We perform Monte Carlo simulations using the Metropolis algorithm. The polymer chain undergoes Rouse dynamics that consists of local corner-flip or end-flip moves~\cite{Doi1986}. It does not violate mutual avoidance with hard wall. The elementary move consists of selecting a random monomer from the chain and flipping it. If the move results in the adsorption of the monomer on the wall, it is always accepted as a move. The opposite move, i.e., desorption of the monomer from the wall, is chosen with the Boltzmann probability $\eta = \exp(-\Delta E/k_B T)$, where $\Delta E$ is the energy difference between two states. The move involving movement of monomer from one desorbed state to another is always accepted. The time is measured in units of Monte Carlo steps (MCSs). One MCS consists of $N$ flip attempts, which means that on average, every monomer is given a chance to flip. Throughout the simulation, the detailed balance is always satisfied and the algorithm is ergodic in nature. It is always possible, from any starting polymer configuration, to reach any other configuration by using the above moves. Before taking any measurements, we let the simulation run for $400\pi/\omega$ in lower frequency regime, and $4000\pi/\omega$ in higher frequency regime, to achieve the stationary state. 
We report quantities in the dimensionless units. The quantities having dimensions of energy are measured in units of $\epsilon$, and the quantities that have dimensions of length are measured in terms of the lattice constant $a$. 
We have taken $k_B = 1$, $\epsilon = 1$ and $a = 1$. 

The displacement, $x(t)$, of the end monomer of the polymer from the wall is the response to the oscillating force $g(t)$. The displacement $x(t)$ is monitored as a function of time $t$ for various force amplitudes $g_0$ and frequency $\omega$. 
The time averaging of $x(t)$ over a complete period,
\begin{equation}
   Q = \frac{\omega} {2\pi}\*\oint x(t)\*dt, 
   \label{eq:chap2Q}
\end{equation}
can be used as a dynamical order parameter~\cite{Chakrabarti1999}.
From the time series $x(t)$, we can obtain the extension $x(g)$ as a function of force $g$. On averaging it over many cycles of the periodic force, we can obtain the average extension 
$\langle x(g) \rangle$ as a function of $g$.
The average extension, $\langle x(g) \rangle$, depends on the frequency $\omega$ of the oscillating force. At higher frequencies, the system does not get enough time to get equilibrated and therefore the $\langle x(g) \rangle$ is not same for the forward and backward paths. This results in a hysteresis loop in force-extension plane. 
The area of hysteresis loop, $A_{loop}$, which is defined by 
\begin{equation}
    A_{loop} = \oint \langle x(g) \rangle dg,  
    \label{eq:TwoA_loop}
\end{equation} 
depends upon the frequency $\omega$ and the amplitude $g_0$ of the driving force and can serve as an another candidate for the dynamical order parameter. 

\section{Results and Discussions} {\label{sec:Results}}

In this section, we discuss the results obtained in our simulations for both the static and dynamic cases. Let us first discuss the static force case briefly.

\subsection{Static Case ($\boldsymbol{\omega = 0}$)}

In the static limit, this model has been solved analytically using the generating function and exact transfer matrix techniques~\cite{Kapri2005}. We first briefly mention the generating function technique and obtain the exact phase diagrams for all the three cases. We then validate our Monte Carlo simulation results by obtaining the force-distance isotherms for various system sizes and then extract the critical unzipping force values at various temperatures using finite-size scaling. We then compare them with the analytical results.

The directed nature of our model makes it possible to calculate the partition function for the polymer via a recursion relation. The generating function technique can then be used to obtain the phase boundary. In this method, the  singularities of the generating function are determined. The singularity nearest to the origin gives the  phase of the polymer and the phase transition occurs whenever the two singularities cross each other. The method is described as follows:
Let $\mathcal{Z}_n(x)$ represents the partition function, in the fixed-distance ensemble, of a polymer of length $n$ with separation $x$ between the $n$th monomer and the wall. Let each lattice site on one side (say $x < 0$) of the wall has a repulsive potential $V (V >0)$. The soft-wall and the hard-wall are then the limiting cases for $V \rightarrow 0$ and $V \rightarrow \infty$, respectively. In the presence of potential $V$, the recursion relation satisfied by the partition function is given by 
\begin{equation}
        \mathcal{Z}_{n+1}(x) =
    \begin{cases}
      [\mathcal{Z}_n(x + 1) + \mathcal{Z}_n(x - 1)]e^{-\beta V} & \text{for}\hspace{0.1cm} x < 0\\
      \mathcal{Z}_n(x + 1) + \mathcal{Z}_n(x - 1) & \text{for}\hspace{0.1cm} x > 0\\
      [\mathcal{Z}_n(x + 1) + \mathcal{Z}_n(x - 1)]e^{\beta \epsilon} & \text{for}\hspace{0.1cm} x = 0.
    \end{cases} \label{eq:recurel}       
\end{equation}
If the above recursion relation is iterated $N$ times with an initial condition $\mathcal{Z}_0(x) = e^{\beta \epsilon} \delta_{x, 0}$, we get the partition function of a polymer of length $N$. 
The generating function for the partition function $\mathcal{Z}_n(x)$, can be taken to be of the form (ansatz) 
\begin{equation}
\hat{\mathcal{Z}}(z,x) = \sum_n z^n \mathcal{Z}_n(x) = 
     \begin{cases}
       \eta^x(z) K(z) & \text{for}\hspace{0.1cm} x > 0\\ 
        \eta'^{-x}(z) K(z) & \text{for}\hspace{0.1cm} x < 0.   
     \end{cases}
     \label{EqCh2:ansatz}
\end{equation}
When the ansatz Eq.~\eqref{EqCh2:ansatz} is used in the above recursion relation (Eq.~\eqref{eq:recurel}), we obtain $\eta = (1-\sqrt{1-4z^2})/2z$, $\eta' = (1-\sqrt{1-4z^2e^{-2 \beta V}})/2ze^{-\beta V}$ and $K(z) = 1/\{1-[\eta'+\eta]ze^{\beta \epsilon}\}$. The singularities coming from $\eta$ and $\eta'$ are $z_1 = 1/2$ and $z'_1 = 1/2\exp{(-\beta V)}$, respectively, and $K(z)$ has the singularity
\begin{equation}
    z_2 = \frac{1}{2}\sqrt{1-\bigg(1-\frac{2e^{-\beta \epsilon}(1-e^{-\beta V}e^{-\beta \epsilon})}{1+e^{-\beta V}(1-2e^{-\beta \epsilon})} \bigg)^2},
\end{equation}
which depends on both the adsorption energy, $\epsilon$, and the potential $V$. In the large length limit, the relevant partition function in fixed-distance ensemble is approximated as $\mathcal{Z}_N(x) \approx \eta^x(z_2)/z_2^{N+1}$ for $x > 0$, with free energy $\beta F = N\ln{z_2}-x\ln{\eta(z_2)}$. The force needed to maintain the separation $x$ is given by $g = \partial F/ \partial x$. The phase boundary, in a fixed-distance ensemble, is then given by  
\begin{equation}
    g(T) = -k_BT\ln{\eta(z_2)}
    \label{eq:gcfd}.
\end{equation}
The zero force melting takes place at $T_c = \infty$ for the soft-wall and $T_c = \epsilon/\ln{2}$ for the hard-wall case. There is a nonzero $T_c$ for any $V < \infty$.

In the fixed-force ensemble, there is an additional force-dependent singularity, $z_3(\beta g_0) = [2 \cosh{(\beta g_0)}]^{-1}$, which comes from the generating function
\begin{equation}
    \mathcal{G}(z, \beta g_0) =\sum_{n = 0}^{\infty} z^n \sum_x \mathcal{Z}_n(x)e^{\beta g_0 x} .
\end{equation}
The phase boundary 
comes from equating the two singularities $z_2 = z_3$ and is given by
\begin{equation}
     g_c(T) = T\cosh^{-1}{\Bigg[ \Bigg\{1-\bigg(1- \frac{2u(1-vu)}{1+v(1-2u)}\bigg)^2 \Bigg\}^{-1/2}\Bigg]},
 \label{eq:Twogcff}
\end{equation}
with $u = e^{-\beta \epsilon}$, and $v = e^{-\beta V}$. This phase boundary obtained in the fixed-force ensemble (Eq.~\eqref{eq:Twogcff}) is identical to the phase boundary obtained in the the fixed-distance ensemble (Eq.~\ref{eq:gcfd}). In the limits $V\rightarrow 0$ and $V\rightarrow \infty$, the above equation simplifies to the phase boundaries for the soft-wall and hard-wall cases, respectively   
\begin{equation}
    g_c(T) =
    \begin{cases}
      T \tanh^{-1}{[1-e^{-\beta \epsilon}]} & \textnormal{ for soft-wall}\\
      T \tanh^{-1}{[1-2e^{-\beta \epsilon}]} & \textnormal{ for hard-wall}.
    \end{cases}
    \label{eq:Twogcffhswal}
\end{equation}
The phase boundary separating the adsorbed and the unzipped phases for all the three cases: hard-wall ($V\rightarrow \infty$), soft-wall ($V\rightarrow 0$) and the wall separating two different media ($V = 1.0$) are shown in Fig.~\ref{fig:3} by lines. The region below the phase boundary represents the adsorbed phase while above it represents the unzipped phase. From Eq.~\ref{eq:Twogcff}, we obtained the critical force as $g_c( T = 0.5) = 0.6557....$, $ 0.4959....$ and $0.4636....$ for soft-wall, wall separating two media ($V =  1.0$) and hard-wall, respectively, at temperature $T = 0.5$ used in this study.

The Monte Carlo simulations can be used to obtain many other equilibrium properties. We perform Monte Carlo simulations on the model to obtain the force $g$ vs average extension, $\langle x \rangle$, of the end monomer of the polymer from the wall. Every data point in the force-distance isotherms is obtained by equilibrating the system for $2 \times 10^5$ MCSs and then averaged over $10^4$ different realizations.

\figThree
\figTwo

In Fig.~\ref{fig:2}, we have plotted the scaled extension $\langle x \rangle / N$, as a function of constant pulling force $g$ for the polymer of various lengths $N = 128, 256$, and $512$ at $T = 0.5$ obtained by using Monte Carlo simulations for (i) the soft-wall [Fig.~\ref{fig:2}(a)], (ii) the wall separating two different media [Fig.~\ref{fig:2}(c)], and (iii) the hard-wall [Fig.~\ref{fig:2}(e)]. From the figure, we can clearly see the existence of the zipped and the unzipped phases. The polymer is in the zipped phase at lower $g$ values with $\langle x \rangle / N \approx 0$, and in the unzipped phase with $\langle x \rangle / N \approx 1$ when the external pulling force $g$ exceeds a critical value $g_c$. Furthermore, with the increase in the chain length $N$, the transition becomes sharper. In the thermodynamic limit, $N\rightarrow \infty$, it would become a step function at a critical value $g_c$. 

The critical value of the force, $g_c$, is obtained by using the finite-size scaling on polymer lengths $N = 128, 256$, and $512$ 
\begin{equation}
\langle x \rangle  = N^{d} \mathcal{F}\Big( \big(g-g_c\big) N^{\phi} \Big),
\label{eq:Twogcmc}
\end{equation}
where $d$ and $\phi$ are the critical exponents. The data shows a very nice collapse for the values:
\begin{enumerate}
    \item  $g_c = 0.65\pm0.02$, $d = 0.96 \pm 0.05$, and $\phi = 1.0\pm0.02$ for the soft-wall case,
    \item $g_c = 0.49\pm0.02$, $d = 0.93 \pm 0.10$, and $\phi = 1.0\pm0.02$ for the wall separating two different media case, and
    \item $g_c = 0.46\pm0.02$, $d = 0.90 \pm 0.10$, and $\phi = 1.0\pm0.02$ for hard-wall case.
\end{enumerate}
The data-collapse obtained using the above exponents are shown in Figs.~\ref{fig:2}(b), \ref{fig:2}(d), and \ref{fig:2}(f) for the soft-wall, the wall separating two different media, and the hard-wall cases, respectively.
The critical force values are obtained by using the above method at various temperatures are plotted by points in Fig.~\ref{fig:3}. They are found to match the analytical results [Eq.~\ref{eq:Twogcff}] obtained using the generating function technique. Let us now use our Monte Carlo simulations to the dynamic case.  

\subsection{Dynamic Case}

In the previous section, we have seen that for the static force case, the exact results are available due to the generating function technique. However, for the dynamic force case, where the polymer is subjected to an oscillating force, the exact results are not available. We therefore use the Monte Carlo simulations, to study the unzipping of polymer subjected to an oscillatory force. We will discuss results for all the three types of walls. 
The oscillating force $g(t)$ defined in Eq.~\eqref{eq:Twoforce} has both the positive and negative cycles. For the soft-wall and the wall separating two different  media, the polymer can cross the wall and therefore both the positive and negative cycles are used in pulling the polymer. However, for the hard-wall case, the polymer cannot cross it and remains adsorbed on the wall during the negative cycle of the force. Therefore, for the hard-wall case, we take the absolute value of the force $g(t)$ to convert the negative cycles also to positive cycles and define 
\begin{equation}
  g_h(t) = g_0 |\sin(\omega_h t)|, 
 \label{eq:TwoFHW}  
\end{equation}
where $\omega_h $ is the frequency of the force $g_h(t)$. On comparing Eqs.~\eqref{eq:Twoforce} and ~\eqref{eq:TwoFHW}, we see that for same time period, the frequency of the force for the hard-wall case is $\omega_h = \omega /2$. Henceforth, we will remember this and drop the subscript $h$ from both $\omega$ and force $g(t)$.   

\figFour

\subsection*{1. Extension}

The response of an oscillating force $g(t)$ is seen in the extension $x(t)$ of the end monomer of the polymer from the wall. In Fig.~\ref{fig:4}, we have plotted the time variation of the scaled extension, $x(t)/N$, for the polymer of length $N = 128$ as a function of time $t$ along with the time variation of the external force $g(t)$, at force amplitude $g_0=1.2$, for two different frequencies 
$\omega = 1.42\times 10^{-2}$ and $6.28\times 10^{-4}$ for the soft-wall ($V = 0.0$), the wall separating two different media ($V = 0.5$), and the hard-wall ($V = \infty$) at temperature $T = 0.5$. The magnitude of the force rises from $0$ to a peak value of $g_0$, which is greater than the critical force $g_c$ required to unzip the polymer at equilibrium, and then falls back to $0$ in both the positive and negative cycles. The scaled extension $x(t)/N$ follows the force $g(t)$ with a lag and its value depends on the frequency of the oscillating force. 

At a higher frequency, $\omega = 1.42\times 10^{-2}$, the force changes very rapidly and the polymer does not get enough time to relax. Therefore, fewer number of monomers are unzipped from the wall resulting in smaller scaled extension. For the soft-wall case, since the polymer experiences similar environment on either sides of the wall, the scaled extension is almost symmetrical in the positive and the negative cycles of the pulling force $g(t)$ [Fig.~\ref{fig:4}(a)]. However, this is not the case for the wall separating the two different media. In this case, the polymer feels a repulsive potential ($V = 0.5$) on side $x < 0$ and prefers to remain adsorbed on the wall. On the other side of the wall $x > 0$, the potential is $V = 0$. Therefore, the scaled extension is not the same for both the positive and negative cycles of the force [see Fig.~\ref{fig:4}(b)]. For the hard-wall case, the polymer is unable to cross the wall at $x = 0$ and only a positive scaled extension is observed, which is shown in Fig.~\ref{fig:4}(c). When the force is oscillating at a lower frequency, $\omega = 6.28\times 10^{-4}$, the polymer gets enough time to relax and attains a fully stretched configuration in the response to the oscillating force. This results a larger scaled extension [see Figs.~\ref{fig:4}(d) -- \ref{fig:4}(f)]. However, the extension is still smaller for the negative cycle of the force in case of the wall separating the two media because the force pulls only a few monomers from the wall due to the presence of a repulsive potential $V = 0.5$ [see Fig.~\ref{fig:4}(e)]. The time series of extension $x(t)$ accumulated for many different cycles can be used to obtain various quantities.               

\figFive
\figSix

\subsection*{2. Hysteresis loops}

In the preceding subsection, we have seen that the response, $x(t)$, of the system during the rise of the magnitude of the force from $0$ to $g_0$ and fall of the magnitude of the force from $g_0$ to $0$ is not the same in both the positive and the negative cycles of the force. At a given temperature, we can obtain the average extension $\langle x(g) \rangle$ as a function of force $g$ by averaging $x(t)$ over a significant number of cycles. The force-distance isotherms thus obtained show hysteresis loops.

In Fig.~\ref{fig:5}, we have plotted average extension $\langle x(g) \rangle$, averaged over $10^4$ cycles, as a function of force $g$ for a polymer of length $N = 128$ at five different frequencies $\omega = 1.57 \times 10^{-2}$, $1.57 \times 10^{-3}$, $1.57 \times 10^{-4}$, $4.18 \times 10^{-5}$, and $1.57 \times 10^{-5}$ at two force amplitudes $g_0 = 1.0$, and $2.5$ for the soft-wall $(V = 0.0)$, wall separating two different media $(V = 1.0)$, and the hard-wall $(V = \infty)$, respectively. The values of the force amplitude $g_0$ is always chosen higher than the critical force $g_c$ needed to unzip the polymer from the wall. The force-distance isotherms for all the three cases show hysteresis loops of various shapes and sizes. 
At $T = 0.5$, the force amplitude $g_0 = 1.0$ is slightly above the phase-boundary, $g_c  = 0.6557\ldots, 0.4636\ldots$, and $0.4959\ldots$, for the soft-wall, the hard-wall and the wall separating the two different media, respectively, and most of the monomers are adsorbed on the wall. For the soft-wall and the wall separating two different media, the polymer can penetrate the wall to gain the configurational entropy and due to this extra entropy, the stationary state of the polymer for the soft-wall case is an adsorbed state. At any finite temperature, the value of the critical force needed to unzip it from the wall is more than at it is at $T = 0$ (see Eq.~\eqref{eq:Twogcff}). Similarly, for the wall separating the two media, the critical force depends on the strength of the repulsive potential $V$ and for smaller values $V$ the phase diagram shows a reentrance  region at lower temperatures~\cite{Kapri2005}. When the polymer is subjected to an oscillating force a with a higher frequency, $\omega = 1.57 \times 10^{-2}$, the force changes very rapidly and it can only unzip a few monomers from the wall. Therefore, we obtain a small hysteresis loop for all the three cases [Figs.~\ref{fig:5}(a)-\ref{fig:5}(c)]. For the soft-wall case, as the polymer experiences similar environment on both sides of the wall, the loop is divided about $\langle x (g) \rangle = 0$ into two equal and symmetrical parts. 
Whereas, in case of the wall separating two different media having different affinities for the polymer, the fast changing force, during the negative cycle, is unable to pull the monomers from the wall against the repulsive potential $V = 1.0$, in the region $x < 0$. Therefore, during the negative cycle of the force, the extension $\langle x(g) \rangle$ remains $0$ with no hysteresis.
In contrast, for the hard-wall case, at any finite temperature, few monomers of the adsorbed polymer at the free end are unzipped to gain the configurational entropy, and therefore, the stationary state of the polymer is a partially zipped state. Therefore, the average extension $\langle x(g) \rangle$ of the polymer even at $g = 0$ is finite (see Fig.~\ref{fig:5}(c)). 

When the frequency is decreased to a relatively lower value, $\omega = 1.57 \times 10^{-3}$, the polymer gets relatively more time to relax. As a result more number of monomers are separated from the wall and the area of hysteresis loop increases. For the soft-wall case, we get two symmetrical loops for the positive and the negative cycles of the oscillating force [see  Fig.~\ref{fig:5}(a)]. In case of the wall separating two different media, the polymer remains adsorbed on wall with no hysteresis in the negative cycle of the force [see Fig.~\ref{fig:5}(b)]. In the positive cycle the force \textcolor{blue}{loop} looks similar to the loop obtained for the hard-wall case as shown in Fig.~\ref{fig:5}(c). At this frequency, the polymer gets fully unzipped at the maximum force value for the hard-wall and the wall separating two different media, resulting in larger loop area. 
As the frequency is decreased further, the polymer now has ample time to relax and therefore, the isotherms for the forward and backward paths begin to retrace each other at higher an lower force values but with a small hysteresis loop at intermediate forces. The area of the hysteresis loop decreases with the decrease in the frequency.

The force-distance isotherms for the higher force amplitude $g_0 = 2.5$ are shown in Figs.~\ref{fig:5}(d) -- \ref{fig:5}(f). The force amplitude $g_0 = 2.5$ is far far above the phase-boundary for all the three cases. For the soft-wall and the wall separating two different type of media, the stationary state is still the adsorbed polymer on the wall (i.e., $\langle x(g) \rangle = 0$ at $g = 0$). The shape of loops for the soft-wall case are similar to the loops obtained for the lower force amplitude $g_0 = 1.0$ but with larger area (Fig.~\ref{fig:5}(d)).     
In case of the wall separating two different types of the media (Fig.~\ref{fig:5}(e)), the polymer remains adsorbed on the wall during the negative cycle of the force at higher frequencies  $\omega = 1.57 \times 10^{-2}$ and $1.57 \times 10^{-3}$ even for force amplitude $g_0 = 2.5$ resulting no hysteresis loop. During the positive force cycle, the loops are similar to $g_0 = 1.0$ case with slightly higher loop area.   
However, on decreasing the frequency to a value, $\omega = 1.57 \times 10^{-4}$, the polymer gets enough time to relax and at higher force values it can overcome the repulsive potential $V=1.0$ on $x<0$ and explore the region during the negative cycle of the pulling force. This results in a hysteresis loop in the region $x < 0$ (see Fig.~\ref{fig:5}(e)) which was absent for $g_0 = 1.0$ [Fig.~\ref{fig:5}(b)]. The loops thus obtained in the positive and negative cycles are not symmetric. For the hard-wall case, the stationary state at $g_0 = 2.5$ is an unzipped state. This can be seen by larger a value for $\langle x(g) \rangle$ at $g = 0$ at a higher frequency $1.57 \times 10^{-2}$. Under the influence of an oscillating force, at this frequency, few monomers at the anchored end of the the completely stretched polymer gets adsorbed on the hard-wall in the backward cycle and are unzipped in the forward cycle resulting in a smaller hysteresis loop (see Fig.~\ref{fig:5}(f)). On decreasing the frequency, more and more number of monomers will get adsorbed on the wall resulting in hysteresis loops of varying shapes and area. In the next subsection, we will see that this change of the stationary state from a partially zipped state at lower force amplitude $g_0 = 1.0$, to an unzipped state for higher force amplitude $g_0 = 2.5$ will give oscillatory behavior in the loop area.  

\figSeven
\figEight

\subsection*{3. Loop Area} 

In this subsection, we explore the behavior of the hysteresis loop area, $A_{loop}$ (defined by Eq.~\eqref{eq:TwoA_loop}), of the curves discussed in the previous subsection. We determine the area of the hysteresis loops using the trapezoidal method. In this method the abscissa is divided into equally spaced intervals and the area of the curve is then the sum of the trapezoids formed by these intervals. 
In our study, the force $g(t)$ changes as a sine function (Eqs.~\eqref{eq:Twoforce} and ~\eqref{eq:TwoFHW}) resulting in a non-uniformly spaced intervals. To make  evenly spaced intervals, we divide the force interval $g \in [0,g_0]$ into 1000 equal intervals for both the rise and fall of the force, in the positive as well as in the negative cycles, and then interpolate the value of $\langle x(g) \rangle$ at the end points of these intervals using cubic splines from the GNU Scientific Library~\cite{Galassi2009}. On these intervals, the loop area, $A_{loop}$, is then numerically calculated using the trapezoidal rule.  

In Figs.~\ref{fig:6}(a) and ~\ref{fig:6}(b), we have plotted $A_{loop}$ as a function of frequency $\omega$ for the polymer of three different lengths $N = 128, 256$, and $512$ at two different force amplitudes $g_0 = 1.0$ and $2.5$ for the soft-wall case ($V = 0$). We find that the loop area $A_{loop}$ depends non-monotonically on the frequency $\omega$ of the periodic force. At very high frequencies, the area of the loop is almost zero. As the frequency of the pulling force decreases, the loop area started increasing. It reaches a maximum value at a specific frequency $\omega^{*}$, to be called as resonance frequency, and then begins to decrease as the frequency is decreased further. At $\omega^{*}$, the natural frequency of the polymer matches with the frequency of the externally applied force and we obtain the maximum loop area. In the limit $\omega \rightarrow 0$, the loop area $A_{loop} \to 0$. At higher force amplitudes (e.g., $g_0 = 2.5$), loop area shows similar behavior as for $g_0 = 1.0$ but with larger magnitudes. We can also see that the frequency $\omega^{*}$ also depends on the amplitude $g_0$ of the oscillating force.  From these figures, it is obvious that the resonance frequency, $\omega^{*}(g_0)$, at which the area of the hysteresis loop is maximum, depends on length $N$ of the polymer. The frequency $\omega^{*}(g_0)$ decreases as the length of the polymer increases and in the thermodynamic limit $N \rightarrow \infty$, we have $\omega^{*}(g_0) \rightarrow 0$. This suggests that $A_{loop}$ satisfies the scaling of the form,
\begin{equation}
    A_{loop} = N^{\delta}  \mathcal{Y}(\omega N^z),
    \label{eq:TwoAscale}
\end{equation}
with $\delta$ and $z$ as critical exponents. On plotting $A_{loop}/N$ data for various chain lengths as a function of $\omega N$ (i.e., for exponents $\delta = 1.00 \pm 0.05$ and $z = 1.00 \pm 0.05$)
, we obtain a nice collapse. The data collapse for $g_0 = 1.0$ and $g_0 = 2.5$ is shown in Figs.~\ref{fig:6}(c) and \ref{fig:6}(d), respectively. The above scaling (Eq.~\ref{eq:TwoAscale}) with exponents $\delta = 1$ and $z = 1$ implies that the loop area scales as $A_{loop} \sim 1/\omega $ in the high-frequency regime (i,e., $\omega \rightarrow \infty$).

\figNine

The loop area $A_{loop}$ as a function of $\omega$ for the wall separating two different media, with $V = 1$ for $x < 0$, are shown in Figs.~\ref{fig:7}(a) and \ref{fig:7}(b) for two different force amplitudes $g_0 = 1.0$ and $2.5$, respectively. For lower force amplitude, $g_0 = 1.0$, the $A_{loop}$ curve behaves similarly as the soft-wall case  but with slightly lower $A_{loop}$ values at $\omega^{*}$ (see Fig.~\ref{fig:6}(a)). The $A_{loop}$ curves at higher force amplitude $g_0 = 2.5$ (Fig.~\ref{fig:7}(b)) are very different from the soft-wall case (see Fig.~\ref{fig:6}(b)). In the present case a new peak starts emerging on decreasing the frequency from $\omega^{*}$. This new peak appears  because of the hysteresis loops emerging for the negative cycle of the force at lower frequencies for higher force amplitude (see Fig.~\ref{fig:5}(e)). The $A_{loop}$ keeps on increasing till the frequency $\omega^{**}$ (say), where it reaches another maximum, and then decreases on decreasing the frequency further to $\omega \to 0$. From Figs.~\ref{fig:7}(a) and \ref{fig:7}(b), we can see that both $\omega^{*}$ and $\omega^{**}$ decreases as $N$ is increased, and in the thermodynamic limit $N\to 0$, both $\omega^{*} \to 0$ and $\omega^{**} \to 0$. The above finite-size scaling form (Eq.~\ref{eq:TwoAscale}) is applicable for this case also. In Figs.~\ref{fig:7}(c) and \ref{fig:7}(d), we have plotted $A_{loop}/N$ vs $\omega N$ for various chain lengths at force amplitudes $g_0 = 1.0$ and $2.5$, respectively. The nice data collapse obtained for both force amplitudes again implies that $A_{loop} \sim 1/\omega $ in the high-frequency regime. 


The loop area $A_{loop}$ as a function of $\omega$ for the hard-wall ($V = \infty$ for $x<0$) are shown in Figs.~\ref{fig:8}(a) and \ref{fig:8}(b) for two different force amplitudes $g_0 = 1.0$ and $2.5$, respectively. For lower force amplitude, $g_0 = 1.0$, the $A_{loop}$ curve for this case also behaves similarly as the soft-wall, and the wall separating two different type of media cases (see Figs.~\ref{fig:6}(a) and \ref{fig:7}(a)). However, for higher force amplitudes (e.g., $g_0 = 2.5$), the $A_{loop}$ curves show oscillatory behavior in the higher frequency regime. These oscillations are similar to the $A_{loop}$ observed for a homopolymer DNA~\cite{Kapri2014} and a block copolymer DNA~\cite{Yadav2021} subjected to a periodic force. The secondary peaks, which are seen only for the hard-wall case at higher force amplitudes, are possible due to the stationary state of the polymer(unzipped configuration) for the hard-wall case at these amplitudes. For all other cases, the stationary state of the polymer is an adsorbed (or zipped) configuration (see previous subsection). Therefore, whenever the force drops below the critical value during the fall and the rise of the force, few monomers of the polymer at the anchored end get adsorbed on the wall and unzipped giving rise to small loop area. On decreasing the frequency, more number of monomers take part in this zipping and unzipping process resulting in increase in loop area. It is observed that the number of secondary peaks increases as the length $N$ of polymer increases. These secondary peaks are higher Rouse modes, whose frequencies are given by $\omega_p = (2p-1)\pi/2N$, with $p = 1, 2, ...$ as integers. The finite-size scaling form given in Eq.~\eqref{eq:TwoAscale} is applicable for this case too. When $A_{loop}/N$ is plotted as a function of $\omega N$ for various chain lengths a nice data collapse is obtained for force amplitudes $g_0 = 1.0$ and $2.5$. The collapse is shown in Figs.~\ref{fig:8}(c) and \ref{fig:8}(d), respectively. This again implies that $A_{loop} \sim 1/\omega $ in the high-frequency regime.


To obtain the scaling behavior in the low-frequency regime (i.e., $\omega \rightarrow 0$), we have plotted in Fig.~\ref{fig:9}, the area of hysteresis loop, $A_{loop}$, as a function of $ \omega^{\beta} (g_0 - g_c)^{\alpha}$ for the polymer of lengths $N = 128, 256$, and $512$ for (i) the soft-wall ($V = 0.0$) [Fig.~\ref{fig:9}(a)], and (ii) the hard-wall ($ V = \infty$) [Fig.~\ref{fig:9}(c)] at force amplitudes $g_0 = 1.0, 1.5$, and $2.5$ and for the wall separating two different media $(V = 1.0)$ [Fig.~\ref{fig:9}(b)] at force amplitudes $g_0 = 1.0, 1.25$, and $1.5$. In the above expression $g_c$ is the critical force, needed to unzip the polymer adsorbed from the wall for the static force case at temperature $T = 0.5$.
An excellent data collapse is obtained for the values of the exponents $\alpha = 1.00 \pm 0.05$ and $\beta = 1.25 \pm 0.05$ in all the three cases. The values of these exponents are found to be similar to the exponents obtained in the the Monte Carlo simulation studies of a homopolymer DNA~\cite{Kapri2014} and a block copolymer DNA~\cite{Yadav2021} subjected to a periodic force. In both these studies the DNA is also modelled, same as the polymer model studied in this paper, by a directed self-avoiding walk in (1+1)-directions. In a recent study, it was shown by Langevin dynamics simulations of a longer DNA chain in 2-dimensions that these exponents remain the same~\cite{Kapri2021}.       

\section{Conclusions}{\label{sec:Summary}}

In this paper, we studied the dynamic transitions in the unzipping of an adsorbed polymer on a attractive surface ( or the wall) subjected to a periodic force with amplitude $g_0$ and frequency $\omega$ using Monte Carlo simulations.
We observed that the force-distance isotherms show hysteresis loops in all the three cases. The shape and the size of the loop depends on the frequency of the periodic force. For the soft-wall case, as the polymer experiences similar environment on both sides of the wall, the loop is divided about $\langle x (g) \rangle = 0$ axis into two equal and symmetrical parts.  Whereas, in case of the wall separating two different media having different affinities for the polymer, loops obtained in the positive and negative cycles are not symmetric. On the other hand, for the hard-wall case, the hysteresis loops are possible only in the region $x > 0$. The behavior of the loop area, $A_{loop}$, depends on both the frequency and the amplitude of the force. We found that $A_{loop}$ shows nonmonotonic behavior as the frequency is varied keeping the amplitude constant. On increasing the frequency, the loop area first increases, it reaches a maximum at frequency $\omega^{*}(g_0)$, and then decreases on increasing the frequency further. For small force amplitudes, $A_{loop}$ shows only one peak at a resonance frequency $\omega^{*}(g_0)$, and it decreases monotonically on increasing the frequencies for all the three cases. However, for higher force amplitudes, the $A_{loop}$ still shows only one peak for the soft-wall and hard-wall cases, whereas, it shows two peaks of different height for the wall separating two different types of media. Furthermore, secondary peaks are also present at higher frequencies. We found that $A_{loop}$ scales as $1/\omega$ in the higher frequency regime, and as $g_0^{\alpha} \omega^{\beta}$ with exponents $\alpha = 1$ and $\beta = 1.25$ in the lower frequency regime for all the three cases. These exponents are same as that obtained in the earlier Monte Carlo simulation studies of DNA chains~\cite{Kapri2014,Yadav2021} and a Langevin dynamics simulation study of a longer DNA chain~\cite{Kapri2021} in 2-dimensions.

\section*{Acknowledgment}
I thank R. Kapri and S. Kalyan for comments and discussions.

\bibliography{Adsorbed_poly}

\end{document}